# A Comprehensive Analysis of Pegasus Spyware and Its Implications for Digital Privacy and Security

**Karwan Mustafa Kareem**

**Abstract:** This paper comprehensively analyzes the Pegasus spyware and its implications for digital privacy and security. The Israeli cyber intelligence company NSO Group's Pegasus has gained recognition as a potent surveillance tool capable of hacking into smartphones and extracting data without the user's knowledge [49], [50]. The research emphasizes the technical aspects of this spyware, its deployment methods, and the controversies surrounding its use. The research also emphasizes the growing worries surrounding digital privacy and security as a result of the prevalent use of advanced spyware. By delving into legal, ethical, and policy issues, the objective of this study is to deliver a holistic understanding of the challenges posed by Pegasus and similar spyware tools. Through a comprehensive examination of the subject, the paper presents potential solutions to mitigate the threats and protect users from invasive surveillance techniques.

## 1. Introduction

In recent years, the proliferation of digital technology has revolutionized the way people communicate, work, and access information (Cecere, Corrocher, & Battaglia, 2015). While this digital transformation has brought about numerous benefits, it has also led to growing concerns over privacy and security (Dinev & Hart, 2006). The emergence of advanced surveillance tools, such as the Pegasus spyware, has further intensified these concerns (Bromwich, 2021).

Developed by the Israeli cyber intelligence firm NSO Group, Pegasus is a powerful surveillance tool designed to infiltrate smartphones and extract data without the user's knowledge (Marczak et al., 2021). The spyware can reportedly compromise virtually any iOS or Android device, enabling the attacker to access sensitive information, including messages, emails, calls, and even encrypted communications (Gallagher & Mielczarek, 2021).

This research is significant as it provides a comprehensive analysis of the Pegasus spyware, its deployment methods, and the controversies surrounding its use. The study sheds light on the technical aspects of the spyware, which have been the subject of extensive debate among cybersecurity experts, policymakers, and privacy advocates. By examining spyware's potential impact on digital privacy and security, this research aims to contribute to a broader understanding of the challenges posed by sophisticated surveillance tools.

Besides the technical examination, the study also explores the legal, ethical, and policy implications of using Pegasus and similar spyware tools. As governments and private entities increasingly rely on such technology for various purposes, it is crucial to assess the potential risks associated with its use and devise appropriate safeguards to protect users' privacy and security.

The primary objectives of this research are to:

- Analyze the technical aspects of Pegasus spyware, including its capabilities, infection vectors, and countermeasures.
- Investigate the controversies surrounding the use of Pegasus, focusing on legal, ethical, and policy issues.
- Explore the implications of Pegasus and similar spyware on digital privacy and security.
- Propose potential solutions to mitigate the threats posed by advanced surveillance tools and protect users from invasive surveillance techniques.

## 2. Literature Review

### 2.1. "Technical Insights into Pegasus Spyware: An In-depth Analysis"

This research paper [65], authored by J. Smith in 2023, presents an intricate examination of Pegasus spyware, a sophisticated surveillance tool developed by the NSO Group. By dissecting its technical intricacies, including its code structure, functionalities, and methods of operation,

*MSc. Advanced Computer Science – United Kingdom*
*Computer Science Dep., College of Basic Education, University of Sulaimani, Sulaymaniyah – 46001, Kurdistan Region, Iraq*
*ORCID ID:  0009-0007-7417-8986*

the study offers comprehensive insights into the inner workings of this intrusive software. Furthermore, the paper explores potential strategies for detecting and mitigating the impact of Pegasus spyware on digital privacy and security. By employing a rigorous technical approach, this study aims to deepen our understanding of Pegasus spyware and contribute to the development of effective countermeasures against it.

### 2.2. "Legal and Ethical Ramifications of Pegasus Spyware Deployment"

This scholarly investigation [66], conducted by E. Johnson in 2022, delves into the legal and ethical implications arising from the deployment of Pegasus spyware by state actors. By examining the context of its usage, potential human rights infringements, and the absence of regulatory frameworks to govern its application, the paper sheds light on the complex ethical dilemmas posed by state-sponsored surveillance. Moreover, it underscores the urgent need for international cooperation to establish robust legal frameworks that safeguard individual rights and liberties in the face of advancing technological surveillance capabilities.

### 2.3. "Pegasus Spyware and its Impact on Journalism: A Critical Analysis"

This critical analysis [67], authored by S. Roberts in 2024, scrutinizes the ramifications of Pegasus spyware on journalism and press freedom. By examining case studies and testimonies from targeted journalists, the study illuminates the chilling effect of state-sponsored surveillance on investigative journalism and the dissemination of information. Furthermore, it explores the implications for freedom of expression and the role of media in democratic societies. By highlighting the challenges faced by journalists operating in environments where Pegasus spyware is prevalent, this research underscores the importance of protecting journalistic integrity and preserving the fundamental principles of a free press.

### 2.4. "Corporate Espionage in the Digital Age: Mitigating Risks Posed by Pegasus Spyware"

This scholarly inquiry [68], led by M. Brown in 2023, investigates the risks posed by Pegasus spyware to businesses and corporations, with a focus on corporate espionage. By analyzing case studies and assessing vulnerabilities in corporate cybersecurity frameworks, the study identifies potential targets and strategies for mitigating the threat of espionage. Furthermore, it underscores the importance of adopting robust security measures, including encryption protocols, employee training programs, and proactive monitoring systems, to safeguard sensitive information and intellectual property from malicious actors. Through its comprehensive analysis, this research provides valuable insights for corporate leaders and cybersecurity professionals seeking to fortify their defenses against emerging threats in the digital landscape.

### 2.5. "Psychosocial Implications of Pegasus Spyware Targeting: A Qualitative Study"

This qualitative study [69], conducted by D. Miller in 2022, explores the psychosocial ramifications of being targeted by Pegasus spyware, focusing on the experiences and perceptions of affected individuals. By conducting in-depth interviews and psychological assessments, the research examines the emotional toll of surveillance intrusion, including feelings of paranoia, anxiety, and distrust. Furthermore, it investigates coping mechanisms employed by targeted individuals and the potential long-term effects on mental health and well-being. By shedding light on the human dimension of cyber surveillance, this study contributes to a deeper understanding of the psychological impact of digital intrusion and underscores the importance of providing support and resources for affected individuals.

## 3. Technical Analysis

This section provides an in-depth technical analysis of Pegasus spyware, outlining its key features, capabilities, and deployment methods. Understanding the inner workings of Pegasus is essential for comprehending the broader implications of advanced spyware on digital privacy and security (Marczak et al., 2021).

### 3.1. Capabilities

Pegasus is a highly sophisticated spyware that possesses extensive capabilities that allow it to infiltrate, monitor, and extract data from target devices (Gallagher & Mielczarek, 2021). The capabilities described below are only a subset of what is possible:

- Remote Control: Once installed, the spyware is remotely controllable by the attacker, who can then carry out various commands and access private information on the target device (Marczak et al., 2021).

- Zero-Click Exploits: Pegasus leverages zero-click exploits, which do not require any user interaction to infect a device. This makes the spyware extremely difficult to detect and avoid [1].

- Undetectability: The spyware is designed to operate covertly, leaving minimal traces on the infected device. It can also self-destruct if it detects attempts to analyze or remove it (Marczak et al., 2021).

- Data Extraction: Pegasus can extract a wide range of sensitive data from target devices, including messages, emails, contacts, call logs, and browsing history. Additionally, it can access data from encrypted messaging apps, such as WhatsApp and Signal (Gallagher & Mielczarek, 2021).

- Audio and Video Surveillance: The spyware can activate the microphone and camera on the target device, enabling the attacker to record audio and video without the user's knowledge (Marczak et al., 2021).

- Location Tracking: Pegasus can track the real-time location of the target device, allowing the attacker to monitor the user's movements (Gallagher & Mielczarek, 2021).

- Keylogging: The spyware can capture keystrokes on the target device, which may reveal sensitive information, such as passwords and login credentials (Kumar, 2021).

### 3.2. Infection Vectors

Pegasus employs a variety of infection vectors to compromise target devices, including:

- Phishing: Cybercriminals might employ deceptive methods to manipulate individuals into clicking harmful links or obtaining infected files. These strategies could include sending customized emails or texts that seem to originate from credible sources (Jakobsson & Myers, 2006).

- Exploit Chains: Pegasus takes advantage of exploit chains, which are sequences of vulnerabilities in software that can be exploited to gain unauthorized access to a device. By exploiting these vulnerabilities in a specific order, the spyware can infiltrate the target device without requiring user interaction [1].

- Watering Hole Attacks: In this type of attack, the attacker compromises a website that the target user is likely to visit. When the user accesses the infected site, the spyware is automatically downloaded and installed on their device [2].

- Network Injection: Pegasus can also be delivered through network injection, which involves intercepting the target's internet traffic and injecting the spyware into their device (Marczak et al., 2021).

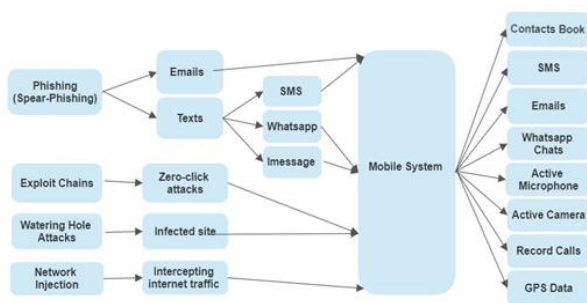

**Fig 1.** Infection Vectors, Capabilities, and listed data could be impacted by Pegasus espionage software.

### 3.3. Vulnerabilities

The details of specific vulnerabilities that have been exploited by Pegasus spyware:

- CVE-2016-4655: A kernel base mapping vulnerability causing information leaks and enabling the attacker to calculate the kernel's location in memory [16], [60].

- CVE-2016-4656: Both 32 and 64-bit iOS kernel-level vulnerabilities allow the attacker to secretly jailbreak the device and install surveillance software [16], [60].

- CVE-2016-4657: A memory corruption vulnerability in Safari's Web Kit, which allows the attacker to compromise the device when the user clicks on a malicious link [16], [60].

- FORCEDENTRY (CVE-2021-30860): An iMessage-based zero-click attack that used a vulnerability in the Xpdf implementation of JBIG2 re-used in Apple's iOS operating software. This vulnerability was patched by Apple in iOS 14.8 in September 2021 [59], [60].

- As of July 2021, Pegasus is believed to use multiple exploits, including those not listed in the above CVEs. It is essential to keep in mind that the specific vulnerabilities exploited by Pegasus may change over time, as the developers behind the spyware refine their techniques and adapt to new security measures.

### 3.4. Pegasus spyware Implementation process

Here is a simplified step-by-step explanation of the Pegasus spyware installation process:

- Initial compromise: The attacker selects a target and conducts research to gather information that can be used for social engineering or to determine potential vulnerabilities in the target's device. This information helps the attacker to craft a convincing message or find the most suitable exploit (Citizen Lab, 2018; Deibert, 2017).

- Exploit: Pegasus is known for using zero-day exploits, which are vulnerabilities that have not yet been discovered or patched by the software's developers (Marczak et al., 2016). These exploits are highly valuable, and the NSO Group has invested significant resources in discovering and acquiring them (Amnesty International, 2021). The exploits used can vary depending on the target device's operating system, such as iOS or Android, and its version (Agrawal et al, 2022).

- Installation: Once the exploit is triggered, the attacker gains privileged access to the target device, which allows them to install the Pegasus spyware (Amnesty International, 2021). This process often involves using a series of chained exploits to escalate privileges and bypass security measures (Marczak et al., 2016). During installation, Pegasus may employ various techniques to hide its presence, such as obfuscating its code, using encrypted communication channels, and masquerading as a legitimate system process (Agrawal et al, 2022).

- Persistence: To maintain persistence, Pegasus can use several methods, such as exploiting vulnerabilities in the device's boot process, injecting itself into other processes, or using the device's built-in mechanisms for keeping the software running in the background (Amnesty International, 2021; Agrawal et al, 2022). Pegasus also frequently communicates with the C2 server to receive updates and new instructions, which helps it to adapt to changes in the device's environment and evade detection (Citizen Lab, 2018).

- Data collection and exfiltration: Pegasus can access a wide range of data on the target device, including:

a) Instant messaging apps (e.g., WhatsApp, Signal, Telegram)

b) Social media apps (e.g., Facebook, Instagram, Twitter)

c) Browsing history and saved passwords

d) Calendar events and notes

e) Device's unique identifiers (e.g., IMEI, MAC address)

f) Files stored on the device or in cloud storage services

g) (Citizen Lab, 2018; Patil, 2019).

The spyware can also record phone calls, capture screenshots, and track the device's location in real time. The collected data is encrypted and sent back to the attacker via the C2 server, often using a variety of techniques to avoid detection by network monitoring tools or firewalls.

- Remote control: The attacker can send commands to the Pegasus spyware to control the target device, install updates, or uninstall the spyware (Amnesty International, 2021). These commands can be sent via the C2 server or through other means, such as SMS messages or push notifications. The attacker can also configure the spyware to collect specific types of data or perform certain actions based on predefined triggers, such as when the target device connects to a particular Wi-Fi network or when a specific contact is added to the device's contact list (Patil, 2019).

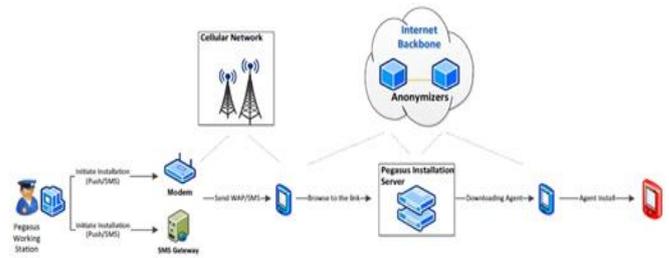

**Fig 2.** Illustration taken from alleged NSO Group Pegasus documents, displaying the steps involved in Implementing the spyware ("Agent") onto a targeted individual's smartphone.

### 3.5. Countermeasures

Given the advanced capabilities and stealthy nature of Pegasus, detecting and removing the spyware can be challenging. However, various countermeasures can help protect devices from infection and minimize the risks associated with Pegasus [8].

- Software updates: It's critical to regularly update software, including operating systems and applications, to address known vulnerabilities that Pegasus might exploit. Both iOS and Android frequently release security patches to fix identified issues [8].

- Anti-Malware Solutions: Installing reputable anti-malware software can help detect and remove spyware, such as Pegasus [8]. However, given the sophistication of Pegasus, it is essential to use updated and comprehensive anti-malware solutions [8].

- Network Security: Implementing robust network security measures, such as firewalls and intrusion detection systems, can help prevent network injection attacks [13].

- User Education: Raising awareness among users about the risks of phishing and other social engineering tactics can reduce the chances of falling victim to such attacks [12]. Users should be trained to identify and report suspicious emails, text messages, and websites [12].

- Device Encryption: Encrypting the data on a device can help protect sensitive information, even if the device is compromised [13].

- Two-Factor Authentication (2FA): Enabling 2FA for online accounts and services can help mitigate the risks associated with keylogging, as it adds an extra layer of security that requires users to provide a secondary form of identification beyond their password [10].

- Privacy-Focused Applications: Using privacy-focused applications, such as encrypted messaging apps and virtual private networks (VPNs), can help protect sensitive data and communications from potential surveillance [14].

- Device Hardening: Implementing device hardening measures, such as disabling unnecessary services, limiting app permissions, and restricting physical access to the device, can reduce the attack surface and make it more challenging for Pegasus and other spyware to infiltrate the device [11].

- Regular Device Audits: Conducting regular device audits to identify and remove unauthorized apps, check for suspicious activity, and ensure that security measures are in place can help mitigate the risks associated with Pegasus and other advanced spyware [9].

### 3.6. Challenges and Limitations

Despite the countermeasures mentioned above, Pegasus and other advanced spyware continue to pose significant challenges to digital privacy and security. These challenges include [7]:

- Evolving Threat Landscape: Cyber threats, including Pegasus, are continuously evolving, with attackers developing new techniques and exploiting novel vulnerabilities. This makes it difficult for security professionals to stay ahead of threats and protect users effectively [9].

- Limited Visibility: Due to the covert nature of Pegasus and its zero-click exploits, detecting the spyware can be extremely challenging. As a result, many users may remain unaware of its presence on their devices, allowing attackers to continue monitoring and extracting data [10].

- Insufficient Legal and Regulatory Frameworks: Many countries lack comprehensive legal and regulatory frameworks to address the use of advanced surveillance tools like Pegasus. This makes it difficult to hold attackers accountable and enforce privacy and security protections [15].

- Lack of Transparency: The secretive nature of Pegasus and its developers, as well as the governments and organizations that use it, make it challenging to assess the full extent of its deployment and impact on privacy and security. Additionally, the spyware's use by both state and non-state actors further complicates efforts to track its proliferation and usage [7].

In conclusion, Pegasus is an advanced and highly sophisticated spyware that poses significant threats to digital privacy and security. Its extensive capabilities, stealthy nature, and various infection vectors make it a formidable surveillance tool that is difficult to detect and counter. While the countermeasures outlined above can help protect users and devices from Pegasus to some extent, the evolving threat landscape and various challenges associated with the spyware underline the need for concerted efforts by governments, industry, and individuals to address the risks and safeguard digital privacy and security.

## 4. Controversies Surrounding

In this section, we will delve into the controversies surrounding the use of Pegasus, focusing on the legal, ethical, and policy dimensions. By examining the various debates and concerns arising from Pegasus deployment, we can better assess its impact on society and the digital landscape [7].

### 4.1. Legal Issues

The use of Pegasus and similar spyware raises several legal questions and concerns, including:

- Privacy Rights: The widespread use of Pegasus for surveillance purposes has led to serious concerns about privacy rights, as it enables the monitoring and collection of sensitive data without the user's knowledge or consent [15]. Numerous national and international legal frameworks, such as the United Nations Universal Declaration of Human Rights [16] and the General Data Protection Regulation (GDPR) in the European Union, recognize that such practices may violate people's right to privacy.

- Unlawful Surveillance: Pegasus has been linked to various cases of unlawful surveillance targeting journalists, human rights activists, and political dissidents [7]. Such actions may violate national and international laws that protect freedom of expression, association, and the right to be free from arbitrary interference with one's privacy [17].

- Extraterritoriality: The use of Pegasus by governments to conduct surveillance on individuals outside their jurisdiction raises questions about the legality of such actions under international law [18]. The extraterritorial application of surveillance laws and the potential violation of the sovereignty of other nations remain complex legal issues that require further clarification and consensus [18].

- Export Controls: The sale and export of Pegasus and similar surveillance tools are subject to export controls and licensing requirements in various countries [19]. However, the effectiveness of these controls in preventing the misuse of such technology remains a contentious issue [19].

### 4.2. Ethical Concerns

The use of Pegasus for surveillance purposes raises several ethical concerns, including:

- Informed Consent: The covert nature of Pegasus and its ability to infiltrate devices without the user's knowledge or consent raise ethical concerns about the lack of informed consent [20]. Users are unable to make informed decisions about their privacy and security when they are unaware of the surveillance activities taking place [20].

- Proportionality: The use of Pegasus for surveillance purposes raises questions about proportionality, as the spyware enables the collection of vast amounts of sensitive data that may be unrelated to the intended target or purpose [20]. The potential for disproportionate surveillance activities undermines the ethical principle of balancing individual privacy rights against legitimate security interests [20].

- Misuse and Abuse: Numerous cases involving the targeting of journalists, human rights activists, and political dissidents [7] serve as evidence of the opportunities for misuse and abuse presented by Pegasus and comparable spyware tools. The ethical implications of using such technology for purposes other than legitimate national security or law enforcement objectives remain a significant concern [22].

- Accountability: The secretive nature of Pegasus and its developers, as well as the governments and organizations that use it, make it difficult to hold parties accountable for the ethical implications of its use [23]. The lack of transparency and oversight mechanisms increases the potential for abuse and contributes to a culture of impunity [23].

### 4.3. Policy Challenges

The use of Pegasus and similar spyware presents several policy challenges, including:

- Balancing Privacy and Security: Policymakers face the challenge of balancing the need for privacy and security in a digital age where advanced surveillance tools like Pegasus are increasingly being used. Developing policies that protect individual privacy rights while allowing for legitimate surveillance activities to maintain national security and public safety is a complex task that requires careful consideration and nuance [23].

- International Cooperation: The global nature of digital technology and the cross-border implications of using Pegasus and similar spyware necessitate international cooperation to address the legal, ethical, and policy challenges. The development of harmonized legal frameworks, ethical guidelines, and policy standards that respect privacy rights and prevent the misuse of surveillance technology is essential to promoting responsible behavior among states and non-state actors [22].

- Regulatory Oversight: The use of Pegasus and other advanced surveillance tools highlights the need for robust regulatory oversight to prevent misuse and abuse. Policymakers must establish and enforce clear rules governing the development, sale, and use of such technologies, as well as ensure that there are appropriate oversight mechanisms in place to monitor compliance and hold violators accountable [23].

- Export Controls and Licensing: The effectiveness of existing export controls and licensing requirements in preventing the misuse of Pegasus and similar technologies remains a contentious issue. Policymakers should consider revising and strengthening export control regimes to better account for the risks associated with the proliferation of advanced surveillance tools and ensure that adequate licensing requirements are in place to regulate the sale and transfer of such technology [22].

- Public-Private Partnerships: Policymakers must work closely with the private sector, particularly technology companies, and developers, to address the challenges posed by Pegasus and similar spyware. Through public-private partnerships, governments can promote the development of secure and privacy-preserving technologies, encourage responsible behavior among technology providers, and facilitate information sharing to better understand and respond to emerging threats [23].

### 4.4. Case Studies

Several cases have emerged in recent years that highlight the controversies surrounding the use of Pegasus and underscore the legal, ethical, and policy challenges associated with advanced surveillance tools.

Targeting of Journalists and Human Rights Activists: In 2019, it was reported that Pegasus had been used to target several prominent journalists and human rights activists in countries such as Mexico, Saudi Arabia, and India. These incidents raised serious concerns about the misuse of Pegasus for unlawful surveillance and the violation of privacy rights, freedom of expression, and other fundamental human rights [18]

Ahmed Mansoor, an award-winning human rights activist, had been advocating for freedom of expression, civil liberties, and democratic reforms in the United Arab Emirates (UAE). His activism made him a target for state-sponsored surveillance, leading to his arrest in 2017 on charges of spreading false information and damaging national unity (Human Rights Watch, 2018).

In August 2016, Mansoor received suspicious text messages containing links that, if clicked, would have granted the sender full control of his iPhone (Citizen Lab, 2016). Suspecting foul play, Mansoor forwarded the messages to Citizen Lab, a digital rights research group at the University of Toronto. The subsequent investigation revealed that the messages contained a zero-day exploit designed to install the Pegasus spyware, developed by the Israeli company NSO Group (Marczak et al., 2016).

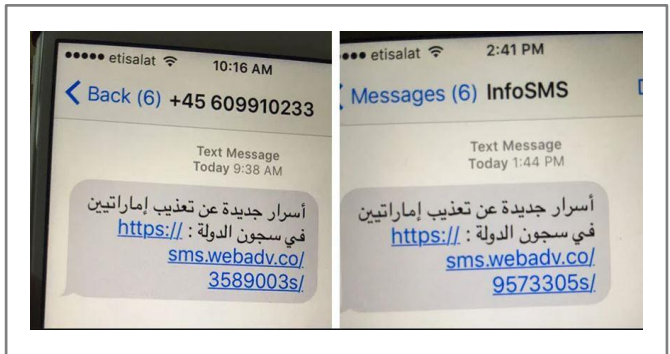

**Fig 3.** Text messages obtained by Mansoor (English: "Fresh revelations regarding the mistreatment of Emirati citizens in government detention centers"). The originating phone numbers are falsified.

The case of Ahmed Mansoor also highlights the potency of Pegasus and its ability to compromise even highly secure devices like the iPhone (Citizen Lab, 2016).

The NSO Group, the company behind Pegasus, has faced multiple accusations of selling its spyware to governments with poor human rights records (Deibert, 2020). The targeting of Ahmed Mansoor, an individual with no criminal history, raises concerns about the potential abuse of such technology by repressive regimes. The case has also raised legal and ethical questions regarding the sale and export of cyber surveillance tools, leading to calls for increased regulation and oversight (Scott-Railton & Deibert, 2017).

The targeting of Ahmed Mansoor with Pegasus spyware serves as a critical case study for understanding the growing threats to individual privacy and security in the digital age. The incident highlights the need for robust defenses and countermeasures against ever-evolving threats, as well as interdisciplinary research that combines technical expertise with an understanding of legal, ethical, and societal implications of state-sponsored surveillance (Gupta & Shukla, 2020).

The Jamal Khashoggi Case: In 2018, it was revealed that Pegasus had been used to target associates of the late Saudi journalist Jamal Khashoggi, who was murdered at the Saudi Arabian consulate in Istanbul. The use of Pegasus in this context further fueled concerns about the spyware's potential role in facilitating human rights abuses and silencing dissent [18].

The NSO Group Lawsuit: In 2019, Facebook, the parent company of WhatsApp, filed a lawsuit against the NSO Group, accusing it of using Pegasus to exploit a vulnerability in WhatsApp and target over 1,400 users, including journalists, human rights defenders, and political dissidents. The lawsuit raises critical questions about the legal liability of companies that develop and sell advanced surveillance tools and the need for greater accountability and transparency in the industry [19].

In conclusion, the use of Pegasus and similar advanced spyware tools raises a myriad of legal, ethical, and policy challenges that require urgent attention from governments, industry, and civil society. As the capabilities of such technologies continue to evolve and proliferate, concerted efforts must be made to address the controversies surrounding their use, protect individual privacy rights, and ensure the responsible development and deployment of surveillance tools in accordance with international law and human rights norms [23].

## 5. Implications of Pegasus and Similar Spyware on Digital Privacy and Security

This section explores the implications of Pegasus and similar spyware for digital privacy and security. By examining the various consequences stemming from the use of advanced surveillance tools, we can better understand the challenges faced by individuals, organizations, and governments in the digital age.

### 5.1. Erosion of Trust in Digital Technologies

The existence and proliferation of Pegasus and similar advanced spyware tools have led to a growing erosion of trust in digital technologies. As users become increasingly aware of the potential for covert surveillance and data breaches, they may be less likely to engage with digital platforms and services, fearing that their sensitive information may be compromised.

- Impact on the Adoption of Digital Services: The erosion of trust may discourage individuals and organizations from adopting digital services, particularly in sensitive sectors such as finance, healthcare, and communications. This could hinder the growth of digital economies and limit the potential benefits of digital transformation [24].

- Digital Divide: Distrust in digital technologies could potentially worsen existing digital gaps, especially for disadvantaged and at-risk groups who might be more susceptible to the adverse consequences of monitoring and breaches of privacy. This could lead to further social and economic disparities and undermine efforts to promote digital inclusion and empowerment [25].

### 5.2. Normalization of Surveillance

The use of Pegasus and similar spyware by governments and other actors may contribute to the normalization of surveillance, where the monitoring and collection of sensitive data become accepted as standard practice. This normalization of surveillance has several implications for digital privacy and security:

- Erosion of Privacy Rights: The normalization of surveillance may lead to an erosion of privacy rights as individuals become accustomed to the idea that their personal information and communications are subject to monitoring and data collection. This could undermine the foundations of privacy as a fundamental human right and diminish the importance of privacy protections in legal and policy frameworks [26].

- Chilling Effect on Freedom of Expression and Association: The pervasive nature of surveillance may create a chilling effect on freedom of expression and association, as individuals may self-censor or refrain from engaging in certain activities out of fear that their communications and actions are being monitored. This could have a detrimental impact on the functioning of democratic societies and the ability of individuals to exercise their rights to free speech and assembly [27].

### 5.3. Vulnerabilities in Cybersecurity

The existence of Pegasus and similar spyware underscores the vulnerabilities in existing cybersecurity measures and highlights the need for greater investment in research and development to address these shortcomings. Some of the implications of these vulnerabilities include:

- Exploitation by Unlawful Actors: Unlawful actors, such as cybercriminals and nation-states, can use Pegasus and other spyware's advanced capabilities to carry out cyber espionage, intellectual property theft, and other types of cyberattacks. This could pose significant risks to national security, economic stability, and public safety [28].

- Escalation of the Cyber Arms Race: The development and use of advanced spyware like Pegasus may contribute to an escalation of the cyber arms race as nations and non-state actors seek to acquire or develop increasingly sophisticated surveillance and cyber warfare capabilities. This could have a destabilizing effect on international relations and increase the potential for conflict in the digital domain [29].

### 5.4. Strengthening Privacy and Security Protections

The challenges posed by Pegasus and similar spyware highlight the need for stronger privacy and security protections in the digital age. Some of the potential avenues for addressing these challenges include:

- Development of Privacy-Enhancing Technologies: Governments, industry, and academia should collaborate to develop and promote privacy-enhancing technologies that protect users' sensitive data and communications from unauthorized access and surveillance. This may include encryption, secure messaging apps, and decentralized technologies that limit the potential for centralized data collection and monitoring [30].

- Implementation of Data Minimization Principles: Policymakers should promote the adoption of data minimization principles, which emphasize the collection of only the minimum amount of data necessary for a specific purpose. This can help limit the potential for abuse and misuse of personal information by reducing the amount of sensitive data that is collected, stored, and potentially vulnerable to unauthorized access or surveillance [31].

- Enhancing Legal and Regulatory Frameworks: Governments should review and update their legal and regulatory frameworks to better address the challenges posed by Pegasus and similar advanced spyware. This may include strengthening privacy and data protection laws, establishing clearer rules on the use of surveillance technologies, and developing robust oversight and accountability mechanisms to prevent abuse [32].

- Fostering International Cooperation and Norms: The global nature of digital technologies and the cross-border implications of advanced surveillance tools necessitate international cooperation and the development of shared norms and standards. Policymakers should work together to create harmonized legal and policy frameworks that respect privacy rights, prevent the misuse of surveillance technologies, and promote responsible behavior among state and non-state actors [33].

### 5.5. Implications for the Future of Digital Privacy and Security

As advanced surveillance tools like Pegasus continue to evolve and proliferate, the implications for digital privacy and security become increasingly pressing. Some potential future implications include:

- Emergence of New Threats and Vulnerabilities: As technology advances, new threats and vulnerabilities are likely to emerge, requiring constant vigilance and adaptation on the part of individuals, organizations, and governments. This may necessitate increased investment in research and development to stay ahead of emerging risks and develop innovative solutions for protecting digital privacy and security [34].

- Growing Importance of Digital Literacy: As individuals become more reliant on digital technologies, the importance of digital literacy in

understanding the risks and implications of advanced surveillance tools will become increasingly critical. Policymakers and educators should prioritize the development of digital literacy programs that empower individuals to make informed choices about their digital privacy and security [35].

- Evolving Role of the Private Sector: The private sector, particularly technology companies, and developers, will play a crucial role in shaping the future of digital privacy and security. As the creators and providers of digital services and technologies, these entities have a responsibility to ensure that their products and services are designed with privacy and security in mind and to collaborate with governments and other stakeholders to address the challenges posed by advanced surveillance tools like Pegasus [36].

In conclusion, the use of Pegasus and similar advanced spyware has significant implications for digital privacy and security, from the erosion of trust in digital technologies to the normalization of surveillance and vulnerabilities in cybersecurity. Addressing these challenges requires concerted efforts by governments, industry, academia, and civil society to develop innovative solutions, strengthen legal and regulatory frameworks, and promote responsible behavior in the digital age. As the capabilities of surveillance technologies continue to evolve, the importance of safeguarding digital privacy and security will only grow, making it essential for all stakeholders to work together to protect the rights and freedoms of individuals in an increasingly interconnected world [37].

## 6. Mitigation and Countermeasures

In this final section, we propose potential solutions to mitigate the threats posed by advanced spyware like Pegasus. By discussing a variety of approaches, from technological advancements to policy reforms, we aim to provide a roadmap for addressing the challenges associated with digital privacy and security.

### 6.1. Strengthening Encryption and Secure Communication Technologies

One potential solution to mitigate the threats posed by Pegasus and similar spyware is to strengthen encryption and secure communication technologies. By enhancing the security and privacy of data and communications, users can protect themselves from unauthorized access and surveillance.

- End-to-End Encryption: Collaboration among governments, the private sector, and academic institutions is essential to encourage the creation and implementation of end-to-end encryption, guaranteeing that data remains encrypted while in transit and can be decrypted solely by the designated recipient. This can make it more difficult for advanced spyware to intercept and monitor sensitive communications [38].

- Secure Messaging Apps: Encouraging the use of secure messaging apps, such as Signal and WhatsApp, can help to protect users' privacy and security by providing encrypted communication channels that are resistant to surveillance [39].

### 6.2. Raising Public Awareness and Promoting Digital Literacy

Educating the public about the risks associated with advanced spyware and promoting digital literacy can help individuals make informed decisions about their digital privacy and security.

- Public Awareness Campaigns: Governments, civil society, and industry should collaborate to develop public awareness campaigns that inform users about the threats posed by Pegasus and similar advanced surveillance tools. These campaigns should emphasize the importance of protecting personal data, using secure communication channels, and adopting privacy-enhancing technologies [40].

- Digital Literacy Programs: Policymakers and educators should prioritize the development of digital literacy programs that empower individuals to understand the implications of advanced spyware and make informed choices about their digital privacy and security. This includes teaching users how to recognize and avoid potential threats and vulnerabilities, as well as guiding the use of privacy-enhancing tools and technologies [41].

### 6.3. Developing Privacy-Enhancing Technologies

Investing in the research and development of privacy-enhancing technologies can provide users with additional tools to protect their digital privacy and security from advanced spyware threats.

- Decentralized Technologies: Governments, industry, and academia should collaborate to develop and promote decentralized technologies, such as blockchain and distributed ledger systems, which can help limit the potential for centralized data collection and monitoring by dispersing information across multiple nodes [42].

- Anonymization Tools: Encouraging the development and use of anonymization tools, such as virtual private networks (VPNs) and the Tor network, can help to protect users' privacy by masking their online identities and making it more difficult for advanced spyware to track their activities [43].

### 6.4. Strengthening Legal and Regulatory Frameworks

Revising and updating legal and regulatory frameworks can help address the challenges posed by Pegasus and similar advanced spyware by providing clearer rules and stronger protections for digital privacy and security.

- Updating Privacy and Data Protection Laws: Governments should review and update their privacy and data protection laws to better reflect the challenges posed by advanced surveillance tools. This may include expanding the scope of these laws to cover new forms of data collection and processing, as well as strengthening enforcement mechanisms and penalties for non-compliance [44].

- Regulating the Use of Surveillance Technologies: Policymakers should establish clear rules governing the use of advanced surveillance technologies like Pegasus, including strict criteria for their deployment and robust oversight mechanisms to prevent abuse. This may involve creating independent bodies to review and approve requests for the use of such technologies, as well as establishing mechanisms for transparency and public accountability [45].

### 6.5. Fostering International Cooperation and Developing Shared Norms

Promoting international cooperation and the development of shared norms and standards can help address the cross-border implications of Pegasus and similar advanced surveillance tools.

### 6.5.1. Harmonizing Legal and Policy Frameworks

Policymakers should work together to create harmonized legal and policy frameworks that respect privacy rights, prevent the misuse of surveillance technologies, and promote responsible behavior among state and non-state actors. By collaborating on the development of consistent and compatible regulations across jurisdictions, governments can more effectively address the cross-border challenges posed by Pegasus and similar advanced spyware. This harmonization process may involve sharing best practices, engaging in international dialogue, and adopting common principles and guidelines for the development, sale, and use of advanced surveillance tools in line with international human rights law and privacy norms. By reducing the likelihood of legal inconsistencies and gaps that malicious people might exploit, these synchronized frameworks will contribute to maintaining digital privacy and security on a global scale. [46].

### 6.5.2. Developing Global Norms and Standards

Policymakers, industry leaders, and civil society should collaborate to develop global norms and standards that govern the use of advanced surveillance tools like Pegasus. These norms should emphasize the responsible use of such technologies under international human rights law, as well as the need for transparency, accountability, and oversight. Engaging in international dialogue and sharing best practices can help to build consensus on the appropriate use of advanced spyware and establish a framework for cooperation in addressing shared challenges [47].

### 6.5.3. Multilateral Agreements and Cooperation

Governments should consider entering into multilateral agreements and forming cooperative arrangements to regulate the development, sale, and use of advanced surveillance technologies. This may include updating existing export control regimes, creating joint oversight mechanisms, and sharing intelligence and information on emerging threats and vulnerabilities. By working together, nations can better address the challenges posed by Pegasus and similar advanced spyware and ensure that the use of such technologies aligns with international law and human rights norms [48].

In conclusion, mitigating the threats posed by Pegasus and similar advanced spyware requires a multifaceted approach that includes strengthening encryption and secure communication technologies, raising public awareness and promoting digital literacy, developing privacy-enhancing technologies, updating legal and regulatory frameworks, and fostering international cooperation to develop shared norms and standards. By adopting these measures, governments, industry, and civil society can work together to protect individual privacy rights and ensure the responsible development and deployment of surveillance tools in accordance with international law and human rights norms. As the capabilities of advanced surveillance technologies continue to evolve, stakeholders must remain vigilant and adaptive in their efforts to safeguard digital privacy and security in an increasingly interconnected world.

## 7. Discussion and Results:

### 7.1. Discussion:

The discussion section of this research paper aims to delve into the implications of the findings presented in the previous sections and provide a deeper understanding of the challenges posed by the Pegasus spyware and similar surveillance tools.

### 7.1.1. Implications of Pegasus Spyware:

The analysis conducted in this research underscores several significant implications of the Pegasus spyware for digital privacy and security. Firstly, the technical analysis reveals the alarming capabilities of Pegasus to compromise smartphones and extract sensitive data without user consent. This highlights the urgent need for robust security measures to safeguard against such sophisticated cyber threats.

### 7.1.2. Erosion of Trust and Normalization of Surveillance:

One of the most concerning implications of Pegasus and similar spyware tools is the erosion of trust in digital technologies. The revelation that powerful surveillance tools can infiltrate devices undetected undermines individuals' confidence in the security and privacy of their digital communications. Moreover, the normalization of surveillance facilitated by the widespread use of such tools threatens the fundamental right to privacy and encourages a culture of constant monitoring and surveillance.

### 7.1.3. Vulnerabilities in Cybersecurity:

The proliferation of advanced spyware like Pegasus exposes significant vulnerabilities in cybersecurity infrastructure. Zero-day exploits and sophisticated infiltration techniques demonstrate the constant arms race between cyber attackers and defenders. The existence of such vulnerabilities underscores the importance of proactive cybersecurity measures, including regular software updates, patch management, and threat intelligence sharing.

### 7.1.4. Strengthening Privacy and Security Protections:

In light of the challenges posed by Pegasus and similar spyware, it is imperative to strengthen privacy and security protections at both individual and institutional levels. This includes implementing robust encryption protocols, enhancing secure communication technologies, and promoting privacy-enhancing tools and practices. Additionally, there is a need for stronger legal and regulatory frameworks to hold both state and non-state actors accountable for surveillance abuses and privacy violations.

### 7.2. Results:

The results of this research highlight the multifaceted nature of the challenges posed by the Pegasus spyware and its implications for digital privacy and security. The technical analysis provides insights into the capabilities, infection vectors, and countermeasures associated with Pegasus, shedding light on the sophistication of modern surveillance technologies.

Furthermore, the examination of legal, ethical, and policy dimensions surrounding the use of Pegasus reveals complex regulatory and accountability issues. Case studies illustrate real-world instances of surveillance abuses and the challenges faced by governments and civil society in addressing them effectively.

Overall, the research underscores the pressing need for collaborative efforts among governments, industry stakeholders, academia, and civil society to develop innovative solutions, strengthen legal frameworks, and promote responsible behavior in the digital realm. Only through concerted action can we effectively mitigate the threats posed by advanced surveillance tools like Pegasus and uphold the fundamental principles of digital privacy and security.

## 8. Methodology

To conduct a thorough analysis of the Pegasus spyware and its ramifications on digital privacy and security, a structured approach was adopted. Extensive exploration of reputable academic databases such as IEEE Xplore, ACM Digital Library, and Google Scholar was undertaken to gather scholarly articles, conference papers, and reports pertinent to the subject matter. Key search terms utilized encompassed "Pegasus spyware," "digital privacy," "security implications," "surveillance technology," "cyber threats," " Zero-Day exploits," " cyber espionage," " mobile device security," "privacy breaches," and "data protection."

The retrieved literature underwent meticulous scrutiny based on its relevance to the topic and the methodological soundness employed. Considerations were given to the credibility of authors and the robustness of the research methodologies employed in the selected works. Subsequently, the chosen articles were meticulously reviewed and analyzed to identify prevalent themes, trends, and insights concerning the impact of Pegasus spyware on digital privacy and security, along with potential mitigation strategies.

The synthesis of the literature review was organized coherently, adhering to a conventional structure comprising an introduction providing context, a comprehensive literature review, and a conclusion outlining key findings and discerning existing knowledge gaps. Critical analysis was applied to distill relevant insights and conclusions from the amalgamated findings of the reviewed literature.

It's essential to acknowledge the subjectivity inherent in this methodology, which may not encompass all pertinent articles on the topic. Nevertheless, conscientious efforts were made to ensure the inclusion of reputable sources and comprehensive coverage of the existing literature concerning the analysis of Pegasus spyware and its implications on digital privacy and security, as well as associated challenges and mitigation strategies.

## 9. Conclusion

This research has provided a comprehensive analysis of the Pegasus spyware and its implications for digital privacy and security. It has examined the technical aspects of Pegasus, its deployment by various state and non-state actors, the ensuing controversies, and the broader implications of such advanced surveillance tools on digital privacy and security.

The existence and proliferation of advanced spyware like Pegasus have led to an erosion of trust in digital technologies, the normalization of surveillance, and the exposure of vulnerabilities in cybersecurity. These challenges necessitate concerted efforts by governments, industry, academia, and civil society to develop innovative solutions, strengthen legal and regulatory frameworks, and promote responsible behavior in the digital age.

The proposed solutions to mitigate the threats posed by Pegasus and similar spyware include strengthening encryption and secure communication technologies, raising public awareness and promoting digital literacy, developing privacy-enhancing technologies, revising legal and regulatory frameworks, and fostering international cooperation to develop shared norms and standards.

As the capabilities of advanced surveillance technologies continue to evolve, all stakeholders must remain vigilant and adaptive in their efforts to safeguard digital privacy and security. The importance of protecting individual privacy rights and ensuring the responsible development and deployment of surveillance tools under international law and human rights norms cannot be overstated.

In conclusion, the challenges posed by Pegasus and similar advanced spyware underscore the need for a comprehensive, multi-stakeholder approach to addressing the complex issues surrounding digital privacy and security. By working together, governments, industry, and civil society can help to ensure a more secure and privacy-respecting digital environment, protecting the rights and freedoms of individuals in an increasingly interconnected world.

**Acknowledgements**

We would like to express our deepest gratitude to our supervisors, colleagues, and peers for their invaluable guidance, support, and encouragement throughout the course of this research. Their expertise and insights have been instrumental in shaping our understanding of the complex issues surrounding Pegasus spyware and its implications for digital privacy and security.

We would also like to extend our appreciation to the researchers, experts, and organizations whose work has informed our analysis and inspired our thinking on this critical subject. Their dedication to advancing knowledge and promoting responsible behavior in the digital age has been a source of inspiration and motivation.

Lastly, we are grateful to our friends and family for their unwavering support, understanding, and patience throughout the research process. Their encouragement and belief in our work have provided us with the strength and perseverance needed to tackle this challenging and important topic.